\documentclass[aps,prl,showpacs,twocolumn]{revtex4}

\def\be{\begin{equation}}
\def\ee{\end{equation}}
\def\beq{\begin{eqnarray}}
\def\eeq{\end{eqnarray}}

\usepackage{amsmath}
\usepackage{amssymb}

\providecommand{\ket}[1]{\lvert #1 \rangle}

\usepackage{graphicx,color}
\usepackage{subfigure}
\usepackage{epstopdf}

\begin{document}
\title{Tripartite non-separability in classical optics}
\date{\today}

\author{W. F. Balthazar$^{1,2}$, C. E. R. Souza$^{2}$, D. P. Caetano$^{3}$, E. F. Galv\~ao$^{2}$, J. A. O. Huguenin$^{1,2}$, and A. Z. Khoury$^{2}$ \footnote{Corresponding author. }}

\affiliation{$^1$Instituto de Ci\^encias Exatas, Universidade Federal Fluminense, 27213-145 Volta Redonda - RJ, Brazil}
\affiliation{$^2$Instituto de F\'\i sica, Universidade Federal Fluminense, 24210-346 Niter\'oi - RJ, Brazil.}
\affiliation{$^3$Escola de Engenharia Industrial Metal\'urgica, Universidade Federal Fluminense, 27255-125 Volta Redonda - RJ, Brazil}

\begin{abstract}

It is possible to prepare classical optical beams which cannot be characterized by a tensor product of vectors describing each of their degrees of freedom. Here we report the experimental creation of such a non-separable, tripartite GHZ-like state of path, polarization and transverse modes of a classical laser beam. We use a Mach-Zehnder interferometer with an additional mirror and other optical elements to perform measurements that violate Mermin's inequality. This demonstration of a classical optical analogue of tripartite entanglement paves the path to novel optical applications inspired by multipartite quantum information protocols.

\end{abstract}

\maketitle

A composite quantum system is said to be entangled when it is not fully described by the state of its components \cite{Schroedinger1935}. Besides indicating a departure from classical physics, entangled states represent an important resource for a number of quantum information protocols \cite{NielsenChuang}. In classical optics, the mode structure associated with different degrees of freedom of the wave field can also be described by complex vector spaces. As examples, an arbitrary polarization can be written as a complex superposition of circularly polarized beams, and the spatial configuration of a paraxial beam can be decomposed in terms of Laguerre-Gaussian beams. These degrees of freedom can be represented on two independent Poincar\'e spheres \cite{Padgett}, in complete analogy with the Bloch sphere used to represent qubit states \cite{NielsenChuang}. Intriguingly, also in classical optics there are field configurations which cannot be described as a tensor product of definite modes of each individual degree of freedom of the system \cite{Spreeuw}. These non-separable structures display a classical analogue of quantum entanglement \cite{Simon,Toppel,Ghose,Pereira}. One example are vector vortex beams, which are non-separable superpositions of transverse modes and polarization states of a laser beam \cite{Quian, Holleczek,Aiello}. This analogy was used to demonstrate the topological phase acquired by entangled states evolving under local unitary operations \cite{Souza}. Recently, it has attracted a growing interest due both to the fundamental aspects involved, but also for potential applications to classical optical information processing \cite{cardano, milione, milione2,forbes,forbes2,shg,rafsanjani,szameit}. Nonseparable structures have also proved their utility in the quantum optical domain \cite{imagepol,quplate, chen2009, barreiro2010, karimi2010,cryptosteve, cryptouff, cryptolorenzo,cnotsteve, cnotuff,environ,khourymilman}. Analogously to its quantum counterpart, classical entanglement has been characterized via the violation of Bell-like inequalities \cite{Borges,Kagalwala,Eberly}. 

Composite quantum systems may have more than two parts. For tripartite systems, Mermin \cite{Mermin} simplified an earlier argument by Greenberger, Horne and Zeilinger \cite{GHZ}, to show that any local hidden-variable theory for tripartite systems must satisfy
\begin{equation}
M= \left<ZZZ\right>-\left<ZXX\right>-\left<XZX\right>-\left<XXZ\right> \leq 2,
\label{kappa}
\end{equation}
where $Z,Y,Z$ represent the Pauli operators. This inequality is violated by the so-called GHZ-Mermin state:
\begin{equation}
\left|GHZ\right>=\frac{1}{2}\left(\left|000\right>-\left|011\right>-\left|101\right>-\left|110\right> \right),
\label{ghz2}
\end{equation}
for which $\left<ZZZ\right>=+1$ and $\left<ZXX\right>=\left<XZX\right>=\left<XXZ\right>=-1$, resulting in $M = 4$, the maximum algebraic violation of Mermin's inequality (\ref{kappa}). In \cite{spreeuw2}, Spreeuw proposed a scheme in which the polarization and propagation paths of a two independent lasers could be used to construct a classical GHZ-like state.

In this Letter we report the experimental preparation and characterization of a non-separable tripartite state in classical optics. By manipulating the path, polarization, and transverse mode degrees of freedom of a laser beam, we prepared a classically non-separable tripartite structure analogous to the GHZ state of Eq. (\ref{ghz2}). We have also devised optical circuits which allow us to perform the measurements required to evaluate the expectation values in Mermin's inequality, and found that it is violated. Our results open new possibilities for the study of entanglement in the classical optical domain as well as for new optical applications inspired by multipartite quantum information protocols.

Our experiments explore three degrees of freedom of a laser beam: its direction of propagation, or {\it path} $(p)$, {\it polarization} $(P)$ and {\it transverse mode} $(M)$. In order to highlight the analogy with quantum states, we will follow reference \cite{Kagalwala} and use Dirac notation for the vector spaces describing these degrees of freedom. The laser beam can be split so that it can be put in an arbitrary superposition of two different paths (denoted by path basis states $\ket{0}_p, \ket{1}_p$). The beam's polarization is described by superpositions of horizontal ($\ket{0}_P$) and vertical ($\ket{1}_P$) polarizations. Finally, we describe the two-dimensional space of first-order Hermite-Gaussian transverse modes as superpositions of basis states $\ket{0}_M \equiv\ket{HG_{01}}$ and $\ket{1}_M \equiv \ket{HG_{10}}$. Any separable pure state of this tripartite system corresponds to a tensor product of well-defined states of the three degrees of freedom: $\ket{p}\otimes \ket{P} \otimes \ket{M} = \ket{pPM}$.

\begin{figure}
     \centering
     \includegraphics[scale=0.7,trim=1.5cm 4.8cm 2cm 3cm, clip=true]{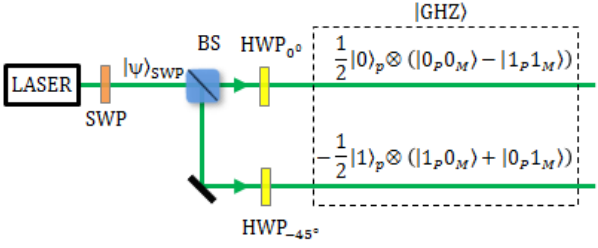}
     \caption{Experimental preparation of a classical analogue of the $GHZ$ state. $SWP$ stands for s-waveplate, $BS$ stands for beam splitter, $HWP$ stands for half wave plate.}
     \label{Fig1}
\end{figure}

In Fig. \ref{Fig1} we show the optical circuit we used to prepare the laser beam state analogous to the GHZ state of Eq. (\ref{ghz2}). A vertically polarized laser beam passes through a {\it s-waveplate} $(SWP)$ to generate a radially polarized beam \cite{vectorvortex}, described by the non-separable polarization/transverse mode state $\ket{\psi_{swp}}=\frac{1}{\sqrt{2}}\left(\ket{0_P0_M}+\ket{1_P1_M}\right)$. The beam is then split in two by a $50/50$ beam splitter $(BS)$, which introduces our dichotomic path degree of freedom. We now operate separately on the two path components. State $\ket{0}_p \otimes \ket{\psi_{swp}}$ goes through a half-wave plate rotated by an angle of $0^\circ$ with respect to the horizontal, resulting in state $\frac{1}{\sqrt{2}}\ket{0}_p\otimes\left(\ket{0_P0_M}-\ket{1_P1_M}\right)$. The other path component $\ket{1}_p \otimes \ket{\psi_{swp}}$ passes through the half-wave plate rotated by an angle of $-45^\circ$ with respect to the horizontal, resulting in state $-\frac{1}{\sqrt{2}}\ket{1}_p\otimes\left(\ket{0_P1_M}-\ket{1_P0_M}\right)$. The overall beam state comprising the two paths then corresponds to an optical analogue of state $\ket{GHZ}$ of Eq. (\ref{ghz2}). Note that the $\ket{PM}$ state on path $\ket{0}_p$ is an even parity eigenstate, whereas the other path carries an odd parity eigenstate.

\begin{figure}[h!]
     \centering
     \includegraphics[scale=0.6,trim=4cm 3.7cm 4cm 3cm, clip=true]{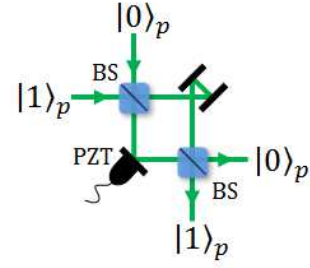}
     \caption{The Mach-Zehnder interferometer with one additional mirror ($MZIM$). The input and output ports are represented by path states $\ket{0}_p$ and $\ket{1}_p$ as in the Figure. $PZT$ stands for piezoelectric transducer.}
     \label{Fig2}
\end{figure}

\begin{figure}
     \centering
     \includegraphics[scale=0.7,trim=2.9cm 2.5cm 5cm 
     2cm, clip=true]{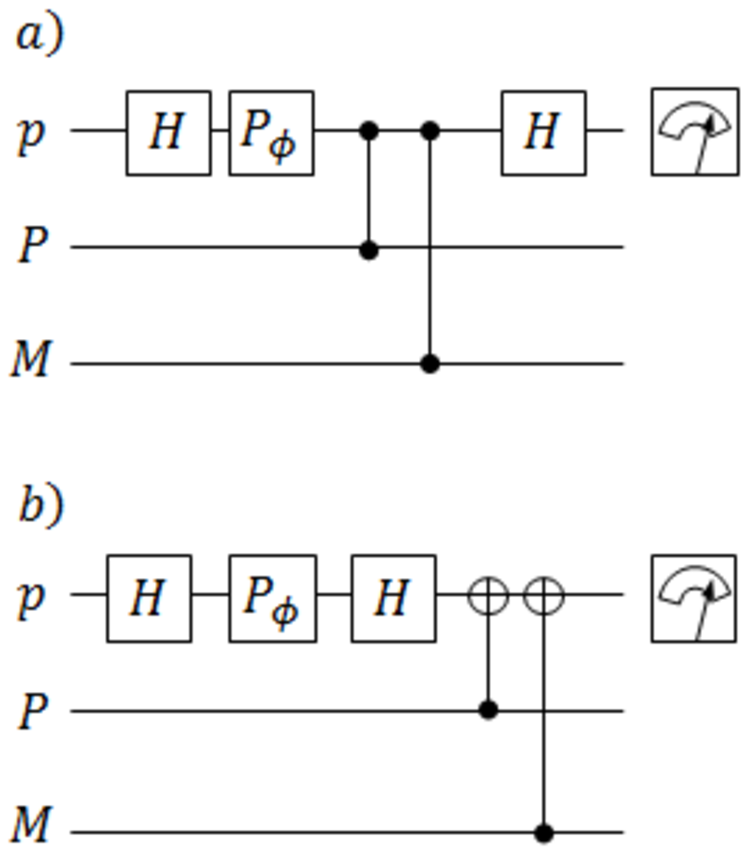}
     \caption{Three-qubit circuits that correspond to the action of $MZIM$ on the three degrees of freedom of a laser beam. a) Direct translation of the $MZIM$'s action on the three degrees of freedom; b) Simpler, equivalent circuit obtained with the circuit identities described in the main text.}
     \label{Fig3}
\end{figure}

Our key experimental tool, represented in Fig. \ref{Fig2}, is a Mach-Zehnder interferometer with an additional mirror $(MZIM)$ \cite{mzim}, with a phase difference $\phi$ between the two arms which can be tuned using a piezoelectric transducer ($PZT$).  To understand how the $MZIM$ works,  let us find the circuit whose action on three qubits corresponds to the interferometer's action on our three degrees of freedom. The first beam-splitter implements a Hadamard gate on the path state $\ket{p}$. Then we have a relative phase shift, modelled in our circuit by a phase gate $P_\phi=diag(1, e^{i\phi})$. This is followed by a double reflection in arm $\ket{1}_p$, which introduces a $-1$ phase on both polarization and transverse modes, but only if the path is $\ket{1}_p$; this is modelled by two gates $CZ=diag(1,1,1,-1)$, controlled by the path $\ket{p}$ and acting on $\ket{P}$ and $\ket{M}$. Finally, the second beam-splitter implements a second Hadamard gate on the path $\ket{p}$. We thus see that the $MZIM$'s action can be directly mapped to the quantum circuit in Fig. \ref{Fig3} a).

We now use the two circuit identities $H^2=1$, $H_i CZ_{i,j} H_i=CNOT_{j\to i}$ on the circuit in Fig. \ref{Fig3} a) to obtain the simpler equivalent circuit in Fig. \ref{Fig3} b). The initial gate sequence $H P_{\phi} H$ on $\ket{p}$ represents the action of a simple Mach-Zehnder interferometer, in which the phase difference $\phi$ can be adjusted to give any chosen beam-splitting ratio at the output modes. The double mirror, modelled by the $CNOT$ gates, exchanges the outgoing intensities of output paths $\{ \ket{0}_p, \ket{1}_p\}$ if the parity of $\ket{PM}$ is odd.

We start by calibrating the $MZIM$ with phase difference $\phi=0$. As we have already described (see Fig. \ref{Fig1}), our GHZ-like state conveniently consists of parity $\ket{PM}$ eigenstates in each path.  To calibrate the $MZIM$ so that $\phi=0$ we simply block path $\ket{1}_p$, adjusting the $PZT$ so that all the even-parity $\ket{PM}$ beam in input path $\ket{0}_p$ is directed to output path $\ket{0}_p$.

\begin{figure}[h!]
     \centering
     \includegraphics[scale=0.7,trim=0cm 1cm 5.5cm 1cm, clip=true]{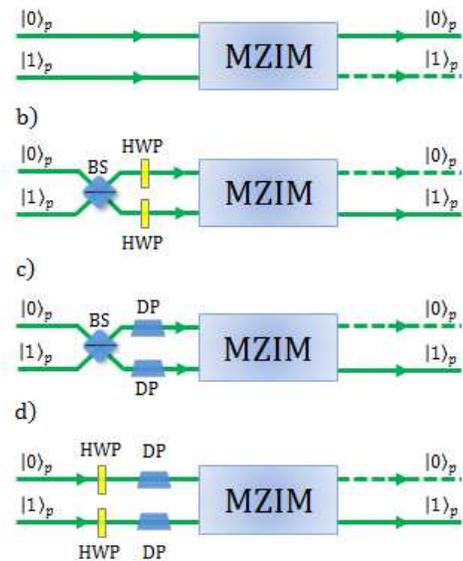}
     \caption{Experimental setup to measure the mean values  $\left<ZZZ\right>$,  $\left<XXZ\right>$, $\left<XZX\right>$,  and  $\left<ZXX\right>$, a)-d), respectively. Dashed lines represent the output where zero intensity is expected.}
     \label{Fig4}
\end{figure}

As we now show, this calibrated $MZIM$ with $\phi=0$ allows us to measure the expectation value $\left<ZZZ\right>$ in Mermin's inequality (\ref{kappa}). Note that even-parity (odd-parity) $\ket{PM}$ states entering via path $\ket{0}_p$ exit via $\ket{0}_p$ ($\ket{1}_p$). Similarly, odd-parity (even-parity) $\ket{PM}$ states entering via path $\ket{1}_p$ exit via $\ket{0}_p$ ($\ket{1}_p$). In summary, we observe that the interferometer's action when $\phi=0$ is to direct the even-parity $\ket{pPM}$ components (for which $\left<ZZZ\right>=+1$) to output path $\ket{0}_p$, and the odd-parity $\ket{pPM}$ components (for which $\left<ZZZ\right>=-1$)  to output path $\ket{p}_1$. In Fig. \ref{Fig4} a) we represent the direct measurement of $\left<ZZZ\right>$ using the $MZIM$. If $I_0$ ($I_1$) is the output intensity at output path $\ket{0}_p$ ($\ket{1}_p$), then the expected value $\left<ZZZ\right> = (I_0- I_1)/(I_0+I_1)$.
We imaged the two output ports of the $MZIM$ in a single frame with a CCD camera, and used the images to estimate the relative output intensities $I_0/(I_0+I_1)$ and $I_1/(I_0+I_1)$. The background intensity of each image (dark noise) was subtracted from the measured ones. 

In order to measure the other required operators in Mermin's inequality (\ref{kappa}), we first note that the Hadamard gate $H$ maps Pauli $X$ operators into Pauli $Z$ operators: $H^{-1}XH=Z$. Hence, a Hadamard gate followed by a physical measurement of $Z$ corresponds to a Pauli $X$ measurement. The Hadamard is implemented on the path degree of freedom by a $50/50$ beam splitter ($BS$); on polarization by a $HWP$ oriented at $22.5^\circ$ with respect to the horizontal; and on the transverse modes by a Dove Prism ($DP$) oriented at $22.5^\circ$  with respect to the horizontal. The measurements of $\left<XXZ\right>$, $\left<XZX\right>$ and $\left<ZXX\right>$ are performed by the circuits sketched in Figs. \ref{Fig4} b), \ref{Fig4} c) and \ref{Fig4} d), respectively, each one involving the application of Hadamard gates to two degrees of freedom. The mean values are calculated from the output path intensities as discussed in the $\left<ZZZ\right>$ case.

It is easy to show that Hadamard gates on two degrees of freedom transforms the GHZ state so that the $MZIM$ input path $\ket{0}_p$ now carries an odd-parity $\ket{PM}$ state, and input path $\ket{1}_p$ carries an even-parity $\ket{PM}$ state. For this reason, measurements of $\left<XXZ\right>$, $\left<XZX\right>$ and $\left<ZXX\right>$ are now expected to result in the full intensity outputting the MZIM at path $\ket{1}_p$, corresponding to an expected value of $-1$ for these observables. The calibration is again done by blocking $MZIM$'s input path $\ket{1}_p$ and using the (now) odd-parity $\ket{PM}$ state of $MZIM$ input path $\ket{0}_p$, adjusting the $PZT$ until we obtain maximum intensity in output path $\ket{1}_p$. Due to the changed parity of the $MZIM$ input paths (with respect to the $ZZZ$ measurement), the calibrated phase difference thus obtained is $\phi=0$, as was the case for the $ZZZ$ measurement.

\begin{figure}[!h]
     \centering
    \includegraphics[scale=0.7,trim=1cm 2cm 3cm 1cm, clip=true]{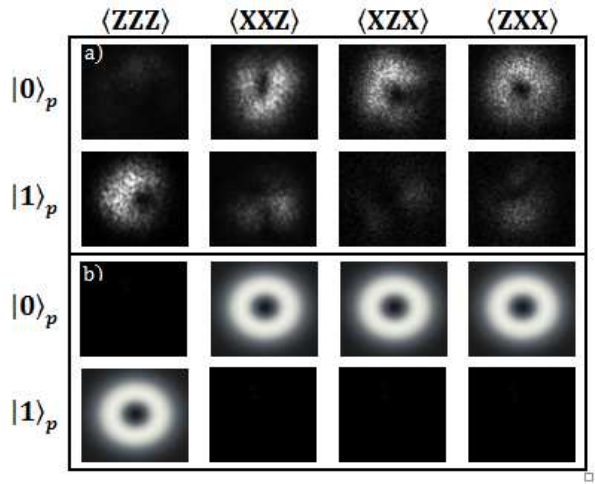}
     \caption{Images at the output paths of the $MZIM$, for all four measurements. a) Experimentally obtained images, whose relative intensity give us the expected values. b) Simulation of ideal experimental outputs.}
     \label{Fig5}
\end{figure}

In Fig. \ref{Fig5} a) we present some experimentally obtained images of output paths $\ket{0}_p$ and $\ket{1}_p$ for all four measurements. As discussed above, in each measurement we ideally expect all the beam's intensity to exit via a single output path. The doughnut shape of the images is due to the radially polarized beam formed in each output path of the GHZ-like structure. In all cases we have experimentally observed some residual intensity coming out of the (ideally) dark output port. This experimental imperfection is more pronounced in the three measurements which require Hadamard gates (to measure $X$ instead of $Z$), as they require a larger number of optical elements than the more direct $ZZZ$ measurement.  These results are in good agreement with the theoretically expected images shown in Fig. \ref{Fig5}b).

\begin{table}[h]
  \centering
  \begin{tabular}{ |c|c|c|c|c|}
   \hline
   \bf $\left<ZZZ\right>$ &  $\left<XXZ\right>$ & $\left<XZX\right>$ & $\left<ZXX\right>$ & $M$ \\
   \hline
      $+0.87 $& $-0.53$ & $-0.63$ & $-0.59$ & $2.62$ \\
       $\pm0.03 $& $\pm0.02$ & $\pm0.02$ & $\pm0.02$ & $\pm 0.05$\\
      \hline
   \end{tabular}
       \caption{Experimentally obtained values for the four measurements involved in Mermin's inequality, and the value of Mermin's quantity $M$. The statistical errors are estimated from the analysis of a number of independently obtained images of each experiment.}
        \label{Tabela1}
\end{table}

From the observed intensities in the two output ports we calculated the expected values of all four operators that appear in Mermin's inequality (\ref{kappa}); they are shown in Table \ref{Tabela1}, together with Mermin's quantity $M$. The error bars represent the statistical variance corresponding to independently obtained experimental images (numbering between $7$ and $9$, depending on the experiment). As can be seen, Mermin's inequality is clearly violated ($M>2$). The less-than-maximal violation can be explained by imperfections in the optical components, e.g. unbalanced $BS$, limited precision in the orientation of $SWP$, $HWP$, and $DP's$, as well as limited sensitivity of the CCD camera ($98 \%$ at $532$ nm). In addition, limited overlap of transverse modes and a fair persistent misalignment in the $MZIM$ contribute to a less-than-perfect violation. Note, however, that these imperfections can only contribute to decrease the amount of violation, as an idealized experiment would be expected to reach the algebraic maximum of $M=4$.

The quantum regime can be reached by attenuating the laser beam to a photon-counting level. In this case, as discussed in Refs. \cite{Borges, Kagalwala}, the prepared states will be genuine tripartite entangled states, where the entanglement is observed among the three degrees of freedom of a single photon.

In summary, we have created tripartite non-separable classical optical states by simultaneously manipulating three degrees of freedom of a laser beam, namely its path, polarization, and transverse mode. We characterised the non-separability by performing measurements that violated Mermin's inequality, an evidence of GHZ-type multipartite non-separability. The measurements combined an $MZIM$ interferometer together with additional optical elements to perform the required optical transformations. The observation of genuine tripartite classical non-separability pushes further this optical analogue of quantum entanglement, opening the path to new optical protocols inspired by multipartite quantum information applications.

\begin{acknowledgments}
We would like to acknowledge the financial support of Brazilian's agencies CAPES, FAPERJ, CNPq, and INCT --- Quantum Information.
\end{acknowledgments}

\end{document}